\documentstyle[11pt,newpasp,twoside,epsf]{article}
\def\reference{\parskip 0pt\par\noindent\hangindent 0.5 truecm}

\def\edcomment#1{\iffalse\marginpar{\raggedright\sl#1\/}\else\relax\fi}
\reversemarginpar

\begin{document}
\title{Hidden Galaxies in the Fornax Cluster}
 \author{M. Drinkwater, M. Waugh, R. Webster}
\affil{School of Physics, University of Melbourne, Victoria
3010, Australia}
\author{D. Barnes}
\affil{Australia Telescope National Facility, PO Box 76, Epping, NSW
1710, Australia}
\author{M. Gregg}
\affil{University of California, Davis, 
       and Institute for Geophysics and Planetary Physics, 
       Lawrence Livermore National Laboratory,
       L-413, Livermore, CA 94550, USA}
\author{S. Phillipps, J.B. Jones}
\affil{Department of Physics, University of Bristol, Tyndall Avenue, Bristol, BS8 1TL, England}

\begin{abstract}
We are using the Multibeam 21cm receiver on the Parkes Telescope
combined with the optical Two degree Field spectrograph (2dF) of the
Anglo-Australian Telescope to obtain the first complete spectroscopic
sample of the Fornax cluster.

In the optical the survey is unique in that all objects (both
``stars'' and ``galaxies'') within our magnitude limits ($16.5\leq B_J
\leq19.7$) are measured, producing the most complete survey of cluster
members irrespective of surface brightness. We have detected two new
classes of high surface brightness dwarf galaxy in the cluster. With
2dF we have discovered a population of very low luminosity
($M_B\approx -12$) objects which are unresolved from the ground and
may be the stripped nuclei of dwarf galaxies; they are unlike any
known galaxies. In a survey of the brighter ($16.5\leq B_J \leq18$)
galaxies with the FLAIR-II spectrograph we have found a number of new
high surface brightness dwarf galaxies and show that the fraction of
star-forming dwarf galaxies in the cluster is about 30\%, about twice
that implied by earlier morphological classifications.

Our radio observations have greatly improved upon the sensitivity of
the standard Multibeam survey by using a new ``basket weave'' scanning
pattern. Our initial analysis shows that we are detecting new cluster
members with HI masses of order $10^8$M$_\odot$ and HI mass-to-light
ratios of 1--2 M$_\odot$/L$_\odot$.
\end{abstract}

\section{Introduction: Selection Effects}

It has long been suggested that optical selection effects limit the
galaxies in optical surveys to a narrow range of surface brightness
(Disney \& Phillipps 1983 and references therein). The idea is that
very low surface brightness (LSB) galaxies are lost in the sky noise
and compact, high surface brightness galaxies would be confused with
stars.  It now seems unlikely that there are large numbers of
undetected giant LSB galaxies (Driver, these proceedings) and the
number of unresolved giant galaxies missed in photographic surveys is
small (Drinkwater et al.\ 1999). However the situation for dwarf
galaxies may be different.

Most flux-limited galaxy surveys are dominated by giant galaxies, so the
only way to study significant samples of dwarf galaxies is to observe
nearby galaxy clusters where the galaxy density is so elevated that
many dwarfs can be detected. The advantage of cluster samples is that
cluster membership can often be assigned on morphological grounds,
avoiding the need for spectroscopy. This has been done very effectively
in surveys of the Virgo and Fornax Clusters which now form the basis
of our knowledge of dwarf galaxies (Binggeli, Sandage \& Tammann 1985;
Ferguson 1989=FCC).  However the lack of spectroscopy becomes a serious
handicap when it comes to unusual types of dwarf galaxy: these may not
be included on morphological grounds.

\begin{figure}
\plotone{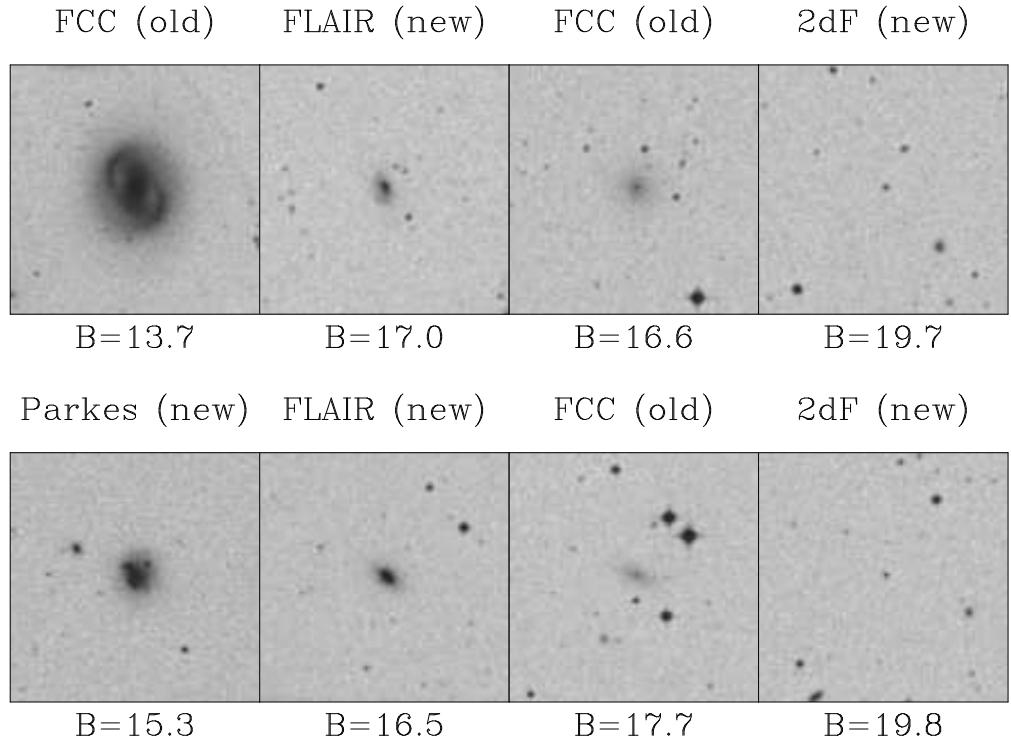}
\caption{New galaxies detected in the Fornax Cluster. These Bj-band
photographic images (from the DSS: see acknowledgements) are all 3 arcminutes
across with North at the top and East to the left.}
\end{figure}

In this paper we present results from an extensive
multi-wavelength {\em spectroscopic} survey of the Fornax Cluster. We
are obtaining the most complete spectroscopic sample of cluster
galaxies ever made, detecting new types of galaxy missed in the
previous morphological surveys. In the optical we are using the Two
degree Field spectrograph (2dF) of the Anglo-Australian Telescope (AAT) to
make a complete survey of faint objects in the core of the cluster and
the FLAIR-II spectrograph of the UK Schmidt Telescope (UKST) to
measure brighter compact galaxies over a six degree field. In the
radio we are making a blind spectroscopic survey of an even larger ten
degree field with the Multibeam 21cm receiver on the Parkes Telescope.
In Fig.~1 we show images of a selection of the cluster galaxies
observed; these are described in more detail below.

\section{``Hidden'' Galaxies Found by Optical Spectroscopy}

\subsection{High Surface Brightness Dwarf Galaxies}

The main difficulty with the morphological studies of dwarf galaxies
in clusters is the possibility that only a subset of cluster members
with familiar properties is selected. In particular, high surface
brightness compact dwarfs may be overlooked because they are
misclassified as giant background galaxies. If such a population were
found it would help explain the evolution of the more compact dwarfs
such as blue compact dwarfs (BCDs) (Drinkwater \& Hardy 1991).

\begin{figure}
\plotone{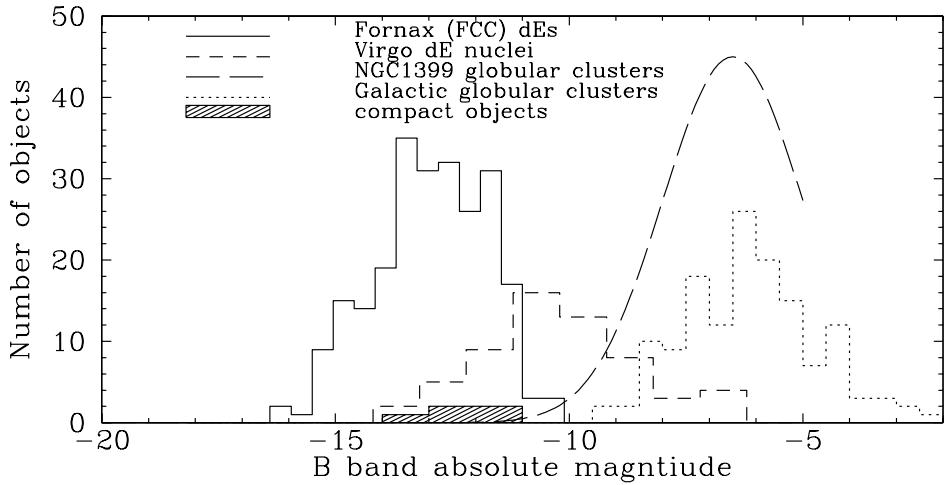}
\caption{Distribution of absolute magnitude of the compact objects
(filled histogram) compared to dEs in the Fornax Cluster (FCC;
solid histogram), nuclei of dE,Ns in the Virgo Cluster
(Binggeli \& Cameron 1991; short dashes), model fit to the globular
clusters around NGC 1399 (Bridges et al.\ 1991; long dashes)
and Galactic globular clusters (Harris 1996; dotted). The
magnitude limit of our survey that found the compact objects
corresponds to $M_B=-11$.}
\end{figure}

We have just completed a spectroscopic search for new compact cluster
dwarfs using FLAIR-II on the UKST (Drinkwater et al.\ in
preparation). We measured 526 brighter ($16.5\leq B_J \leq18$)
galaxies of compact appearance and found ten new cluster members that
were previously classified as background objects. These objects are
all high surface brightness dwarfs, most with strong H-$\alpha$
emission indicative of star formation. Two of these are shown in
Fig.~1 labelled as ``FLAIR (new)'' and are very different to the
``normal'' LSB dwarfs (``FCC (old)'').  We measured spectra of a total
of 108 cluster members including previously known galaxies. These data
showed that star formation as indicated by H-$\alpha$ emission is not
just limited to the dwarf galaxies which were morphologically
classified as late-type, but is also evident in many of the early-type
dwarfs. We find a ratio of star-forming to non-star-forming dwarfs of
1:2 compared to a late-type to early-type ratio of 1:5 according to
the FCC morphological classifications. We also find that the
star-forming galaxies are significantly more extended across the
cluster than those not forming stars actively.

\subsection{Compact Stellar Systems}

We have extended our optical spectroscopy to fainter limits
($16.5<B_J<19.7$) using the 2dF spectrograph on the AAT. This
is almost a
``blind'' survey in that we have observed {\em all} objects both
``stars'' and ``galaxies'' in these limits. The sample (about 3600
objects in the first field) is therefore dominated by foreground
Galactic stars and background field galaxies---but we can afford to
use this complete strategy given the high efficiency of the 2dF system.

\begin{figure}
\plotone{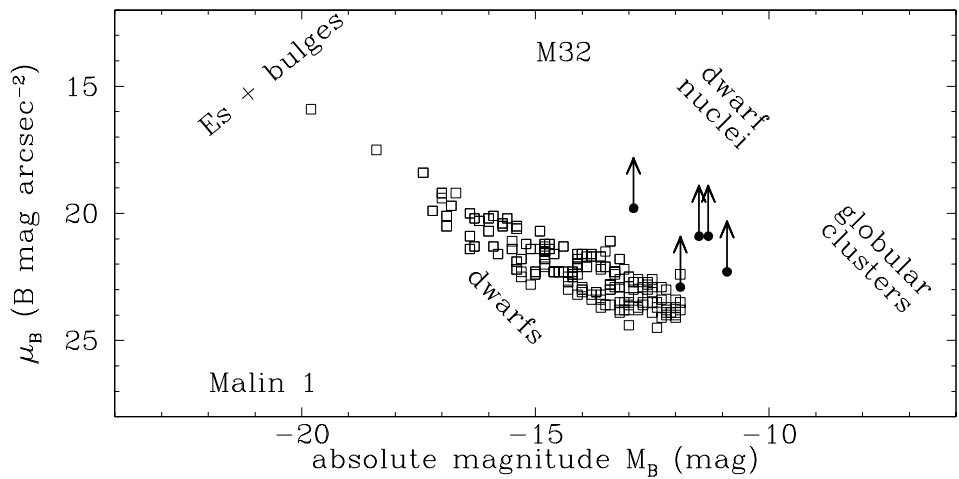}
\caption{Absolute magnitude-surface brightness plane for stellar
systems and subsystems. The squares indicate our measurements of dwarf
galaxies in the Fornax Cluster and the filled circles the new Fornax
compact objects (the surface brightness estimates are lower
limits).  The positions of other populations are from Ferguson \&
Binggeli (1994).}
\end{figure}

Apart from confirming the identification of several ``normal'' dwarf
galaxies in the Fornax cluster we have also discovered a new
population of very compact objects in the centre of the cluster
(Drinkwater et al.\ 2000). These
objects are unresolved in ground-based imaging (see Fig.~1 where they
are labelled as ``2dF (new)'') and have absolute magnitudes of
$-11<M_B<-13$. They have spectra typical of old stellar populations
but they are brighter than any known globular clusters, as is shown in
Fig.~2. The luminosities of these compact objects overlap those of the
fainter known dwarf ellipticals in the cluster, but they are
morphologically distinct, being much more compact. The only objects
they do resemble are the nuclei of nucleated dwarf ellipticals:
perhaps they are the remnants of nucleated dwarfs which have been
stripped in the cluster potential. These objects are unlike any known
type of galaxy or stellar system and occupy a region of the surface
brightness-magnitude plane (Fig.~3) that was previously empty.
We have been awarded {\em Hubble Space Telescope} time to obtain
high-resolution images of these objects to determine what they are.

\begin{figure}
\plotone{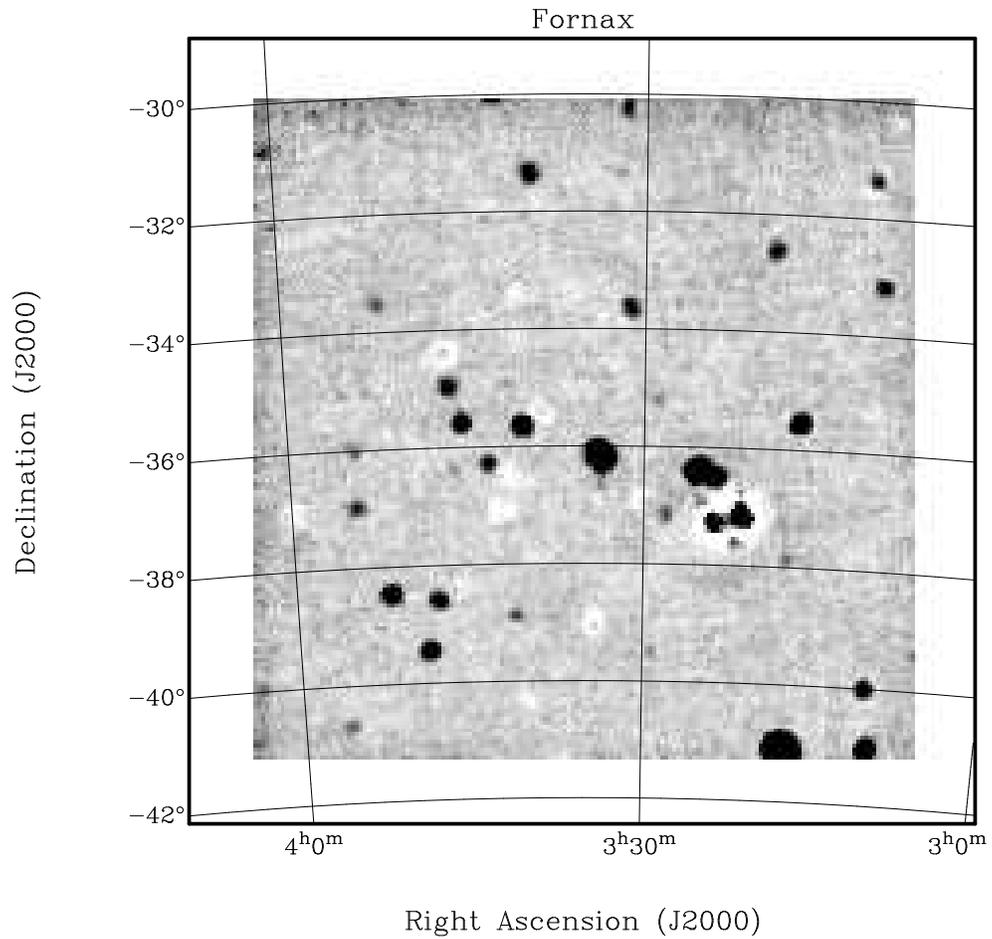}
\caption{Deep basket weave HI map of the Fornax Cluster. The
image measures the maximum signal in the velocity range 400--3000 km/s
at each point in the map. The dark circles are sources. The pale blobs
are caused by continuum sources.}
\end{figure}

\section{``Hidden'' Galaxies Detected in 21 cm Radio Spectroscopy}

In order to extend our survey to larger radii and also to search for
galaxies with high ratios of neutral hydrogen to optical luminosity,
we have made a deep, wide-field survey of the Fornax Cluster with the
Parkes Multibeam receiver. Previous observations with the Parkes
Telescope in ``single beam'' mode (Barnes et al.\ 1997) detected one
new cluster member (``Parkes (new)'' in Fig.~1). This galaxy was
outside the region covered by the FCC.

Our new survey covers an even larger 10 square degree region of sky
centred the Fornax Cluster with 4 times the exposure time of the
standard HI Parkes All-Sky Survey (HIPASS) survey which also covers
this region (see Waugh et al.\ in these proceedings). Unlike the
standard HIPASS survey which uses a single set of North-South scans we
made four sets of scans with half of them in the East-West
direction---hence the term ``basket weave'' we adopt for the new
scanning pattern. We also displaced the field centre by about half the
offset between scans for the second pair of scan sets.  The average
noise per pixel scales with the square root of the exposure time as
expected: about 7 mJy for the basket weave cube compared to 13 mJy for
HIPASS. However the mean and RMS number of scans contributing to each
pixel in the final basket weave cube is $153\pm7$ compared to $42\pm6$
for HIPASS, so the fluctuations in sampling rate from pixel to pixel
have been decreased from 15\% to 5\% making the noise properties much
more uniform. This is shown by comparing the basket weave data in
Fig.~4 with the HIPASS data presented by Waugh et al.\ (these
proceedings).

\begin{figure}
\plotone{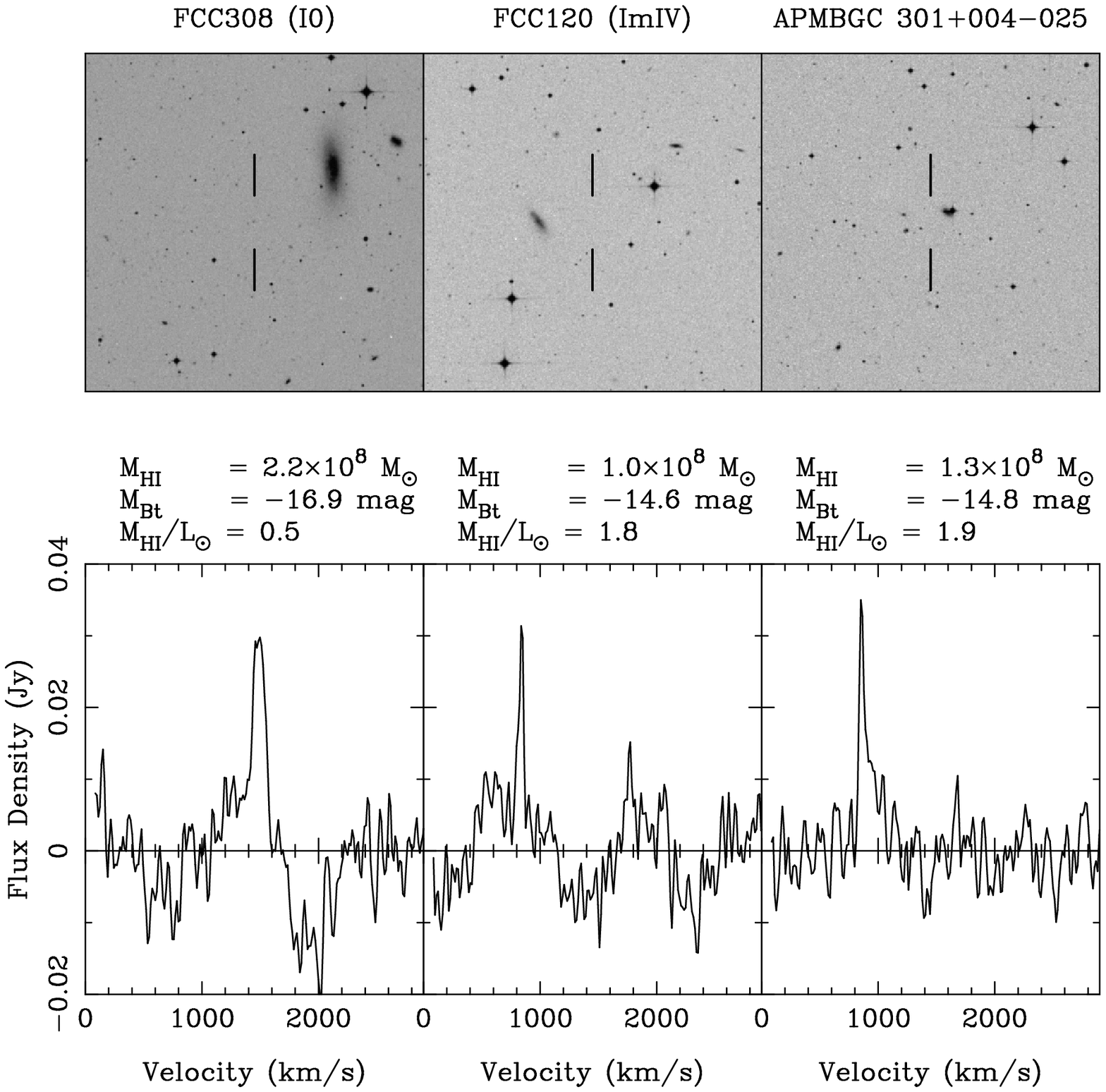}
\caption{Some of the smaller HI sources detected in our basket weave
observations. The upper row shows Bj-band photographic DSS (see
acknowledgements) images centred on the radio detections, all 10
arcminutes across with North at the top and East to the left. The
lower row show the radio spectra of the detections.  }
\end{figure}

We have made a preliminary analysis of the basket weave data, making
strong detections of a number of new galaxies at the 30 mJy level
($M_{HI}\approx1\times10^8$M$_\odot$ for a Fornax distance of 15.4
Mpc) which were not detected in the corresponding HIPASS data. We
present examples of three HI detections in Fig.~5; in each case there
is a probable optical identification. We find no evidence for a large
population of dark neutral hydrogen clouds at the $10^8$M$_\odot$
limit in the Fornax cluster. The first galaxy shown in Fig.~5 (FCC308)
is a relatively bright known cluster member also detected in
HIPASS. The two other galaxies were not detected in HIPASS and are new
cluster members: the first (FCC120) was classified as a cluster member
in the FCC but without a confirmed velocity and the second (APMBGC
301+004-025) was not listed because it was outside the region surveyed
for the FCC.  At this stage it is premature to estimate the total mass
of the cluster in HI, but we do note that these new detections all
have high ratios of HI mass to optical luminosity (see Fig.~5) and
that the cluster HI detections (see Fig.~4) are distributed over a
much larger region than the optical galaxies (core radius of only 0.7
degrees).

\section{Summary}

We have shown that our approach of making complete spectroscopic
surveys has overcome the selection effects evident in previous cluster samples
of dwarf galaxies. The ``new'' members we find are not difficult to
detect optically, but in each case have been excluded from existing
compilations because of their high surface brightness or large
distance from the cluster centre. We have detected unresolved compact
objects, high surface brightness dwarf galaxies and radio
galaxies. When it comes to clusters, galaxies do not have to be
optically faint to be ``hidden'', they just have to be unusual.

Apart from the compact objects, most of the galaxies we found show
evidence of high rates of star formation. In the complete FLAIR-II
sample our data show that star formation is about twice as common in
the dwarf galaxies as would have been implied by the earlier
morphological classifications. Star formation is still important in
the Fornax cluster dwarfs, but not in the cluster core where the
density of star-forming galaxies is reduced. This is consistent with
the radio data which show that the neutral hydrogen in the cluster is
much less centrally concentrated than the optical luminosity.

\acknowledgments

We wish to thank Virginia Kilborn for use of her galaxy detection
software and Ivy Wong for assistance with some of the Parkes
observing.  The Digitized Sky Surveys (DSS) were produced at the Space
Telescope Science Institute under U.S. Government grant NAG
W-2166. The images are based on photographic data
obtained using the UK Schmidt Telescope.


\begin{references}
\reference Barnes D.G., Staveley-Smith L., Webster R. Walsh W. 1997, MNRAS, 288, 307
\reference Binggeli, B.,  Cameron, L.M.,  1991, A\&A, 252, 27
\reference Binggeli, B., Sandage, A., Tammann, G.A., 1985, AJ, 90, 1681
\reference Bridges, T.J., Hanes, D.A., Harris, W.E., 1991, AJ, 101, 469
\reference Disney, M.J., Phillipps, S., 1988, MNRAS, 205, 1253
\reference Drinkwater, M.J., Hardy, E., 1991, AJ, 101, 94
\reference Drinkwater, M.J., Jones, J.B., Gregg, M.D., Phillipps,
S., 2000, PASA, 17, in press
\reference Drinkwater, M.J., Phillipps, S., Gregg, M.D., Parker,
Q.A.,Smith, R.M., Davies, J.I., Jones, J.B., Sadler, E.M., 1999, ApJ,
511, L97
\reference Ferguson, H.C., 1989, AJ, 98, 367 (FCC)
\reference Ferguson, H.C., Binggeli, B., 1994, A\&ARv, 6, 67
\reference Harris, W.E., 1996, AJ, 112, 1487
\end{references}
\end{document}